\documentstyle[pra,aps,12pt]{revtex}
\tighten
\begin{document}
\draft
\title{Barnett-Pegg formalism of angle operators, revivals, and flux lines}
\author{Arul Lakshminarayan}
\address{Physical Research Laboratory,\\
Navarangpura, Ahmedabad,  380 009, India.}   
\maketitle
\newcommand{\newc}{\newcommand}
\newc{\beq}{\begin{equation}}
\newc{\eeq}{\end{equation}}
\newc{\kt}{\rangle}
\newc{\br}{\langle}
\newc{\longra}{\longrightarrow}

\begin{abstract}
We use the Barnett-Pegg formalism of angle operators to study a
rotating particle with and without a flux line. Requiring a finite
dimensional version of the Wigner function to be well defined we find
a natural time quantization that leads to classical maps from which
the arithmetical basis of quantum revivals is seen.  The flux line,
that fundamentally alters the quantum statistics, forces this time
quantum to be increased by a factor of a winding number and determines
the homotopy class of the path.  The value of the flux is restricted to the
rational numbers, a feature that persists in the infinite dimensional
limit.
\end{abstract}
\pacs{03.65.Bz. To appear in Phys. Rev. A}


The multi-valued nature of angle has rendered subtle a quantum
mechanical formulation of angle operators that are conjugate to angular
momentum.  A related problem has been the phase operator of a single
mode of an electromagnetic field, conjugate to the number operator. We
study the effect of one approach on an important model in physics,
namely that of a charged particle rotating around a magnetic flux tube
that does not penetrate the particle path and which 
was shown to be an anyon \cite{Wilczek}. It is appropriate that
such an attempt be made as at the heart of this system is also 
the multi-valued nature of a gauge field. 

The treatment of the angle operators is within the framework proposed
by Barnett and Pegg (BP) \cite{BP}. This relies on the construction of
a finite dimensional Hilbert space in which physical expectation
values are calculated and subsequently the limit of infinite
dimensionality is taken. The periodic boundary condition on the
angular momentum states, however, results in a global or topological
effect that is reflected as flux quantization.

The BP approach is not without its own issues although it has been
applied in various situations. For instance one apparent difficulty
that has been resolved only recently is the definition of an angular
velocity operator \cite{Johal}.  The discussion below is not so much
on the correctness of the methodology but rather observes the
restrictions that are imposed due to the structure of the Hilbert
space in the BP formalism and the way these are transcended in the
infinite dimensional limit. For instance the angular momenta
associated with the composite can take on only rational values while
time takes on a dense set of real values.

We argue this, quite simply, by constructing a Wigner function
in the finite Hilbert space and requiring this construct to be valid
for all time.  Wigner functions for finite quantum mechanics has also
been constructed earlier in \cite{Wootters} while Wigner functions for
photon number and phase have been studied by several authors
\cite{Vaccaro}.  Our following arguments are closely related to, and
informed by, the allowed boundary conditions of quantized torus maps
for which a recent treatment is found in \cite{KMR99}, and the Wigner
function we use has been used in this context previously. Maps are
discrete time (in this context canonical as well) transformations,
while time is continuous for the charged particle rotating the flux
line.  However in the BP formalism the finite dimensionality of the
Hilbert space imposes a natural quantization of time.

Both the classical and the quantum dynamics of a rotating
particle is essentially an unperturbed twist map, quantization 
restricting the dynamics to rational
sub-lattices. Phenomena peculiar to the quantum dynamics of the rotator
such as revivals and fractional revivals due to constructive
interference \cite{BluhmKP} are present almost identically in the
corresponding classical arithmetical dynamics. This is an exact
mathematical rendering of the periodic bunching of (classical)
athletes observed on the racing tracks and used previously as a
pedagogic device for quantum revivals \cite{Nauen}.

We first consider a bare rotating particle with no threading magnetic
flux.  We briefly recapitulate the BP formalism.  A finite dimensional
Hilbert space $H^{2l+1}$ is obtained on imposing a maximum (minimum)
angular momentum quantum number $l (-l)$. The angle states are the
following superposition of the angular momentum states $|m \kt , \; m=
-l, \ldots, l $, which are eigenstates of $\hat{L}_{z}$ with
eigenvalues $m \hbar$:

\beq |\theta_{n} \kt \, =\,
\frac{1}{\sqrt{2l+1}} \sum_{m=-l}^{l} \, \exp(-i m \theta_{n}) \, |m
\kt, 
\eeq 
the angle eigenvalues are 

\beq \theta_{n}\, =\, \frac{2 \pi n}{2 l +1}
\; \; (n=0,1,\ldots, 2l).  
\eeq 

The state $|m=l \kt$ is ``shifted'' by the exponential of the angle
operator, defined from the angle states, to $|m=-l \kt$.  Similarly
the angle state $|n=2l \kt$ is shifted to the zero angle state; thus
the essential Hilbert space structure is that of a quantized torus
phase space. The limiting process is $l \rightarrow \infty$, $\hbar$
fixed and leads to the cylindrical phase space of a rotator. The BP
formalism is in many ways similar to that developed by Schwinger
\cite{Schwing} with different motivations.  The usual limit of quantum
maps on the torus connects $\hbar$ and $l$ and is the classical limit
that leads to a toral phase space. However for any finite value of $l$
and $\hbar$ the Hilbert space structures are identical and we now use
this.

Hannay and Berry \cite{HB} while studying chaotic Anasov mappings of the torus
onto itself used Wigner functions that are ``Dirac delta brushes'' to
describe the effects of the quantum map. Following them, we may define
a Wigner function for the states in $H^{2l+1}$ as \cite{FMR}
\beq
W(\psi, s,r) =  \frac{1}{2(2l+1)}
\, \sum_{k=-l}^{l} c_{k}^{\ast} \, c_{r-k} \,
\exp\left[i \pi s(r-2k)/(2l+1)\right].
\eeq
Here $-2l \, \le\, r \, \le \,2l+1$ and $0 \, \le\, s \, \le \, 4l+1$
are integers and $c_{k}= \br k |\psi \kt $ is the momentum
representation of the wavefunction.  It is known that there is a price
to pay for the definition of the Wigner function in finite dimensional
Hilbert spaces. If $N$ is the dimension of the space the Wigner
function is defined on an $(2 N)^2$ grid rather than a $N^2$ grid,
although there are only $N^2$ independent components, sets of fours
being related to each other.  If $N$ is odd, as in the case of the
rotator, one representative point may be chosen of the four and the
Wigner function can be displayed on an $N^2$ grid.

Thus in the BP formalism the Wigner function is defined over
$\{2(2l+1)\}^2$ points.  The angle is related to $s$, while the
angular momentum to $r$. For instance the angle corresponding to $s$
is $\theta=2 \pi s/(2 (2l+1))$. As usual a sum over $s$ gives the
probability $c_{r/2}^{\ast}c_{r/2}$ if $r$ is even and zero otherwise;
and a sum over $r$ gives the probability density in the position basis
only for the even lattice points and is zero otherwise. For this
property to hold under time evolution the representative $N^2$ grid
must map onto itself.

Now the Hamiltonian for a particle of mass $M$ on a ring of radius $R$ is
\beq
\hat{H} \, =\, \frac{\hat{L}^2_{z}}{2 M R^2}
\eeq
Therefore the time evolution is simply specified by a phase change in the
momentum basis:
\beq
c_{k}(t)\, =\, \exp(-i \tau k^2) c_{k}(0),
\label{ck}
\eeq where $\tau=\hbar t/(2 M R^2)$ is dimensionless time. We have
implicitly assumed that the Hamiltonian  generates time evolution in
the usual manner.  The Wigner function evolution is then
straightforward: 

\begin{eqnarray}
W(\psi(t), s,r)\, =\,\frac{1}{2(2l+1)} 
\sum_{k=-l}^{l} \exp(i \tau k^2) c_{k}^{\ast}(0) \nonumber \\ \times \exp(-i \tau
(r-k)^2) c_{r-k}(0) \exp\left[\frac{i \pi s (r-2k)} {(2l+1)}\right].
\label{Wpsit1}
\end{eqnarray}
Simplifying which we get
\beq
W(\psi(t),s,r)\, =\, W\left(\psi(0), s-\frac{r \,(2l+1) \tau}{\pi}, r\right).
\label{Wpsit2}
\eeq

In the context of linear torus maps such a relationship has been known
for sometime \cite{HB,FMR}.  However note that we have not used
any discrete mapping as yet: time has been a continuous variable. We
are now forced to ``quantize'' time due to the finite dimensionality
of the Hilbert space. Indeed we will have to require that $(2l+1)
\tau/\pi =$ integer (say $2 \,j$, $j$ integer) 
for the Wigner function above to be even
defined. This implies that $t\, =\, j T_{0}$, where
\beq
T_{0}\, =\, \frac{4 \pi M R^2}{(2l+1) \hbar}
\label{T}
\eeq
is the time unit chosen (and not half this).  One reason is that $T_0$
will then correspond to the period of rotation with the highest
allowed  angular momentum ($l \hbar$) in the large $l$ limit, a
natural time scale in finite dimensional spaces. The other has to do
with the representative grid mapping onto itself so that the Wigner
function retains its properties which disallows $T_0/2$ as the
fundamental unit of time.

However, there does not seem to be any a-priori reason why other
integer multiples of $T_0$ should not be the time quantum.  The
fundamental path corresponding to $T_0$ is a single encircling, while
topologically distinct paths on the circle, members of the homotopy
group $\pi_1$ with higher winding numbers, correspond to these other
times.  In fact this ambiguity is essential for the existence of
anyonic behaviour in the flux-tube-charge-particle system. If we were
to restrict, for some reason, the time quantum to $T_0$ even with the
flux, we would end by restricting the composite to be either bosons or
fermions but nothing ``in between''. This is discussed further below.

We show now how this time quantization reveals the simple arithmetical basis
for quantum revivals in the rotator while becoming irrelevant in the
large $l$ limit.  If at time $t=j T_{0}$ the arguments of the Wigner
function are $s_{j}$ and $r_j$ then at time $(j+1) T_{0}$ they are
$s_j-2 \, ~r_j$ and $r_j$. Making the identification
$y_{j}=r_j/(2(2l+1))$ and $x_{j}=\theta_j/(2 \pi) = s_{j}/(2(2l+1))$
we then get the map \beq \begin{array}{ccl} x_{j+1}&=&x_{j}-2 \,y_{j}
\;\; (\mbox{mod}\, 1)\\ y_{j+1}&=&y_{j}.  \end{array} \eeq This simple
unperturbed twist map is an example of a parabolic cat map and the
associated linear transformation $((1,-2),(0,1))$ is of the
quantizable checker-board form of alternating odd and even integer
entries \cite{HB}.

It is easy to see that this simple classical arithmetical map shows
revivals and fractional revivals on rational sub-lattices and that the
time for the full revival $t_{\mbox{rev}}$ then simply corresponds to
$j=N/2=(2l+1)$. Thus $t_{\mbox{rev}}= j T_{0}=4 \pi M R^2/\hbar$;
while at $t_{\mbox{rev}}/2$ there is a fractional revival into two
subsidiary ``wavepackets''. These of course coincide with the usual
revival times of a quantum rotator.  In the final analysis the finite
dimensionality $l$ plays no part, as it has no physical basis; and
neither does the choice of $T_0$ over its integer multiples for the
time quantum.  We must note that the classical fractional revivals are
generally not perfect as this depends on the divisibility properties
of $N$, the imperfections disappearing in the $N \rightarrow \infty$ limit.

The effects of a magnetic flux line on a rotating charged particle
forms the bound Aharanov-Bohm (AB) effect and the composite system is
a model of an anyon \cite{Wilczek}.  As such there are no restrictions
on the flux line threading the particle path. We may either think of the
charged particle as having a new energy spectrum and single-valued
wavefunctions or having the same energy but multi-valued wavefunctions
that acquire a phase as we complete a loop. The differences between
the two and their physical content has been discussed before, for example in
\cite{Silverman}. However if we apply the BP formalism to a particle
with a flux line and demand that the Wigner function remain defined at
all instances of time, we would end in restricting the phase acquired
to either $0$ (periodic) or $\pi$ (anti-periodic), unless the time
quantum is suitably dilated.

The usual minimal substitution leads to the Hamiltonian:
\beq
\hat{H} \, =\, \frac{(\hat{L}_{z}-q \Phi/2 \pi c)^2}{2 M R^2},
\eeq
the charge being $q$ and the speed of light is $c$ while $\Phi$ is 
the threading  magnetic flux. We adopt the first viewpoint and 
consider single-valued momentum eigenstates and an altered energy 
spectrum.
Then
\beq
c_{k}(t)\, =\, \exp(-i \tau (k-\alpha)^2) c_{k}(0),
\label{ckf}
\eeq
where $\alpha=q \Phi/2 \pi \hbar c=\Phi/\Phi_0$. We then derive that 
\beq
W(\psi(t),s,r)\, =\, W\left(\psi(0), s-\frac{(r-2 \alpha)(2l+1) 
\tau}{\pi}, r\right).
\label{Wpsitf}
\eeq

We again require that the Wigner function be defined for all time.  If
the time is still quantized as before in Eq.~(\ref{T}) then we must
require that $4 \, \alpha$  be an integer.  It must be noted
that this quantization of the flux is {\it independent} of the finite
dimensionality $l$ and therefore survives the limit $l \rightarrow
\infty$. This is unlike the quantization of time which takes on a
dense set of real values and transcends to a certain extent the
limitations of a finite Hilbert space. The allowed values of $\alpha$
are then 1 (or 0) and $1/2$. The values $1/4$ and $3/4$ are ruled out
as the representative Wigner lattice is not left invariant for these
choices and time evolution will result in non-zero probabilities at
points off the lattice.  Thus the phase acquired by a $2\, \pi$
rotation must either be $0$ or $\pi$ and the composite is no more a
model of an anyon, as it can have only either bosonic or fermionic
character.

These severe restrictions on the phase are linked then to the choice
of $T_0$ as the time quantum.  However if $\alpha=m/n$, $m$ and $n$
being co-prime integers, we may dilate $T_0$ to $n \, T_0$ and the
restrictions on the flux are lifted and anyonic behaviour is
recovered. Thus for any dimensionality rational flux lines will be
supported. In the infinite dimensional limit, as time acquires a dense
set of values, the flux line is still restricted to the rational
numbers. The time $n \, T_0$ corresponds to $n$ rotations at the
highest allowed angular momentum and appears as the time quantum that
is linked to the flux.  Thus the fundamental motion is not a single
encircling of the flux line, but a path with a winding number
$n$. Thus the homotopy class of the particle path is determined 
by the flux.

We could have adopted the other viewpoint of
multi-valued momentum eigenfunctions and unchanged energy
eigenvalues. Then the Dirac delta brush on which the Wigner function
exits will be supported at points shifted away from the integers by
$\alpha$ and once again will result in identical restrictions.  Thus
if the BP formalism of angle operators is combined with the Wigner
function evolution AB magnetic flux lines in flux-tube-charge-particle
composites are restricted to the rationals and can be interpreted
as corresponding to homotopically distinct paths.

\thanks{ I wish to thank Prof. J. P. Keating for sending
me the work in \cite{KMR99}. I have also profited from discussions 
with Prof. G. S. Agarwal and Dr. J. Banerji}

\end{document}